*Article*

# Software Implementation of Optimized Bicubic Interpolated Scan Conversion in Echocardiography

**Olivier Rukundo [1,*], Samuel Emil Schmidt[1] and Olaf T. Von Ramm[2]**

[1] Department of Health Science and Technology, Aalborg University, Denmark
[2] Department of Biomedical Engineering, Duke University, USA

**Abstract:** This paper introduces a novel approach leveraging objective image quality assessment (IQA) metrics to optimize the outcomes of traditional bicubic (BIC) image interpolation and interpolated scan conversion algorithms. Specifically, feature selection through line chart data visualization and computing the IQA metrics scores are used to estimate the IQA-guided coefficient-*k* that updates the traditional BIC algorithm weighting function. The resulting optimized bicubic (OBIC) algorithm was subjectively and objectively evaluated using natural and ultrasound images. Results showed that the overall performance of the OBIC algorithm was equivalent to 92.22% of 180 occurrences when compared to the BIC algorithm, while it was 57.22% of 180 occurrences when compared to other algorithms. On top of that, the OBIC interpolated scan conversion algorithm generally produced crisper and better contrast cropped ultrasound sectored images than the BIC algorithm, as well as other interpolated scan conversion algorithms mentioned.

## 1. Introduction

Echocardiography is a routine ultrasound imaging, used for diagnosing, managing, and following up on patients with heart disease, providing a comprehensive evaluation of the heart and blood vessels but whose effectiveness still relies on the operator's skill in image acquisition and measurement [1]. About 10% to 15% of routine echocardiograms have poor image quality, [2], which is one key factors that worsens the measurement variability in medical diagnosis. Several attempts have been made to improve ultrasound image quality [3], and ultrasound companies continuously work to improve their echocardiographic systems' image quality [2]. Despite these efforts, ultrasound system device settings, such as gain, focus, depth, dynamic range, level of compression, sector angle, zoom, time gain compensation, and other parameters, remained manually adjustable so that the desired structures were brought into view in the highest quality possible [4]. Other image quality settings and parameters can also affect echocardiography image quality and become performance bottlenecks on any ultrasound imaging systems. One of such settings is the choice of the interpolation algorithm for scan conversion operations. In this study, we introduce an optimized BIC interpolation algorithm (OBIC) for interpolated scan conversion operations to improve cardiac ultrasound image quality. Our hypothesis is to take into account the complexity of the quality content of natural images, extract quality information or coefficient of interest and use it to update the BIC's sixteen coefficients-based weighting function, more details of which are provided in the next part II. Note that software-based interpolated scan conversion operations involve converting a vector image into a raster image, and an interpolation algorithm of interest is used to fill in gaps between available echo-data samples to provide an estimate of the echo reflectivity at each grid point in the image [5], [6], [7], [8]. Also, note that the process of generating such estimates at each grid point must preserve the existing echo-data samples, because image interpolation algorithms, which exist in various categories, by definition, must preserve the existing echo-data samples [9], [10], [11], [12], [13], [14], [15], [18], [28]. It is important to note that current narrow artificial intelligence-based approaches to image interpolation problem [18], produce enlarged images with interesting perceptual quality but without guarantee to respect the fundamental definition of image interpolation. This paper is organized as follows: Part I gives a brief introduction relevant to the topic. Part II is dedicated to materials and methods used. Part III presents the results and discussions of the outcome. Part IV gives the conclusion on the outcome and future.

## 2. Materials and Methods

### *2.1. Scan conversion*

The scan conversion is the process of converting a vector image into a raster image by transforming polar coordinate ultrasound data into Cartesian coordinate data visualizable by standard digital monitor systems. Figure 1 illustrates Eq. (1) and Eq. (2) for translating from polar coordinates to Cartesian coordinates which constitutes the basics for scan conversion.

$$x = r \times cos(\theta) \quad (1)$$

$$y = r \times sin(\theta) \quad (2)$$

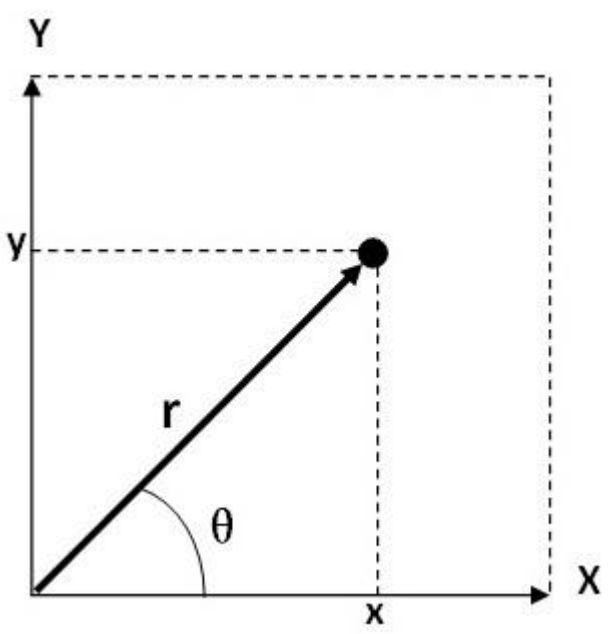

**Figure 1**: Polar to Cartesian

In the past, ultrasonic scan converters were implemented using custom hardware, digital signal processors (DSPs), and field-programmable gate arrays [19], [29]. Today, with the advances in digital sensor technology, it has become possible to implement scan converters in software platforms [20], [21]. Figure 2 shows an example of bilinear interpolation scan conversion (with variables used in the scan conversion operation in typical ultrasound systems) provided by Texas Instruments (TI) for TI C64x+™ based DSP devices.

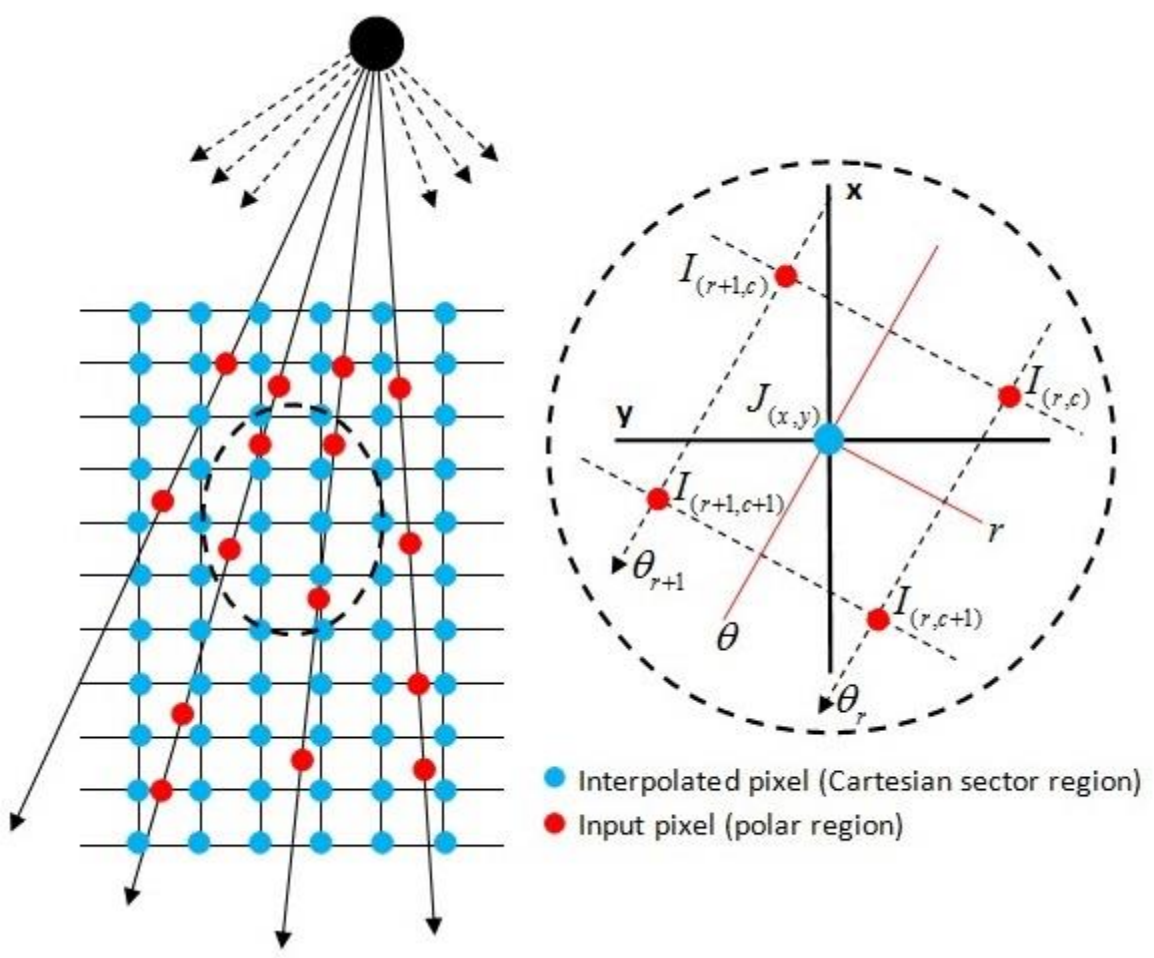

**Figure 2**: Diagram of bilinear interpolated scan conversion

In Figure 2, $J_{(x,y)}$represents interpolated data. $\theta$ and $r$ represent the angular and radius information, respectively. $I_{(:,:)}$represents scan conversion input samples. In interpolated scan conversion, it is important to note that input to the scan converter is scanned echo data and the output is typically data that needs to be displayed on a digital monitor [20]. Also, it is important to note that the probe collects A-mode vectors, received at fixed, angular intervals, from all the specified angles, to form a B-mode frame. The B-mode frame is simply the two-dimensional array formed by storing all the A-mode vectors, from all the angles, in consecutive order. In brief, the B-mode frame is fed into the scan converters to form a sectored image [19].

## *2.2. Image interpolation algorihms*

### 2.2.1. Nearest neighbor interpolation algorithm

Nearest neighbor interpolation algorithm (NN) is a non-extra pixel category of interpolation algorithms and the most efficient [9],[16],[26],[30]. Figure 3 illustrates a simplified example of the destination image, showing four grid coordinates or locations $(r,c)$ , $(r,c+1)$, $(r+1,c)$ and $(r+1,c+1)$ supposedly filled in with source image pixels and five empty locations to be filled in. Note that the output function $D(r',c')$ (see Eq. 3) can take the value of the pixel nearest to $\left(\frac{r\prime}{L},\frac{c\prime}{L}\right)$ from the source image's coordinates by rounding the new coordinates $\left(\frac{r\prime}{L},\frac{c\prime}{L}\right)$ obtained thanks to the Eq. 4 and Eq. 5 of the source image function $S(r,c)$. Here, the ceil rounding function [16] is very important because it decides which nearest neighbor to pick from the source image before copying it in the destination image.

$$D(r',c') = S(\lceil(r)\rceil, \lceil(c)\rceil) \quad (3)$$

$$r = r' \times \left(\frac{1}{L}\right) \quad (4)$$

$$c = c' \times \left(\frac{1}{L}\right) \quad (5)$$

Note that the destination coordinates along the $x$-axis are represented by $r'$ while the coordinates along the $y$-axis are represented by $c'$ and the scaling ratio between the source and destination images is denoted by L.

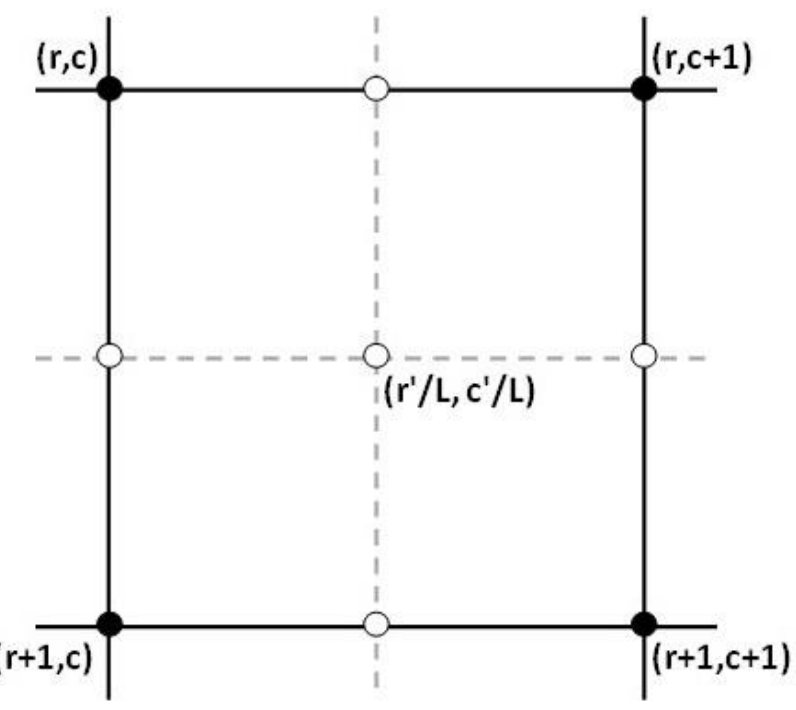

**Figure 3**: Illustration of the pixel grid for NN image interpolation

Also, note that the source coordinates along $x$-axis are presented by $r$ while the source coordinates along $y$-axis are represented by $c$. For the development of the NN interpolated scan conversion algorithm, Eq. 4 and Eq. 5 were adapted to B-mode frame from an ultrasound scanner of interest following examples in [19], [20].

### 2.2.2. Bilinear interpolation algorithm

Bilinear interpolation algorithm (BI) is an extra pixel category of interpolation algorithms known for acceptable effectiveness in comparison with the NN algorithm [5], [10],[12],[13],[14],[15],[24],[26],[30]. The BI's output function $D(r',c')$ is based on Eq. 6 's weighting scheme that it uses to estimate the extra pixel value for use in the destination image. It is important to note that, in Eq. 6, $a$ and $b$ are weighting variables defining distances from grid coordinates of interest in x- and y-directions in the destination image.

$$
\begin{aligned}
D(r', c') = {} & (1-a) \times (1-b) \times S(r,c) + \\
& (a) \times (1-b) \times S(r, c+1) + \\
& (1-a) \times (b) \times S(r+1, c) + \\
& (a) \times (b) \times S(r+1, c+1) \quad (6)
\end{aligned}
$$

For the development of the BI interpolated scan conversion algorithm, Eq. 6 was adapted to B-mode frame from the ultrasound scanner following examples in [19], [20].

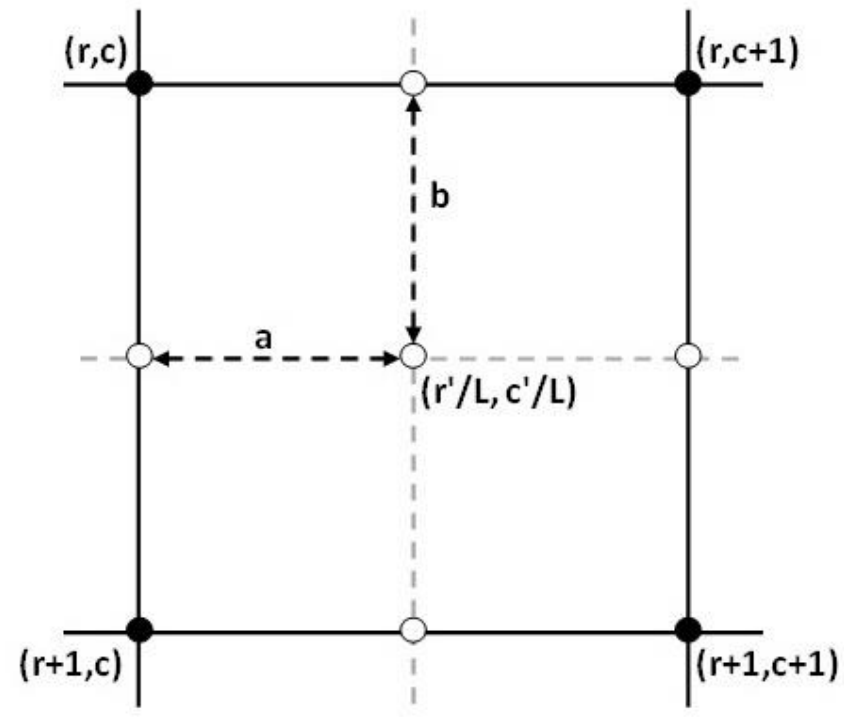

**Figure 4**: Illustration of the pixel grid for BI image interpolation

2.2.3. Bicubic interpolation algorithm

Bicubic interpolation algorithm (BIC) is another extra pixel category of interpolation algorithms known for being more effective compared to both NN and BI algorithms [26] [28]. The BIC's output function D(r',c') is based on Eq. 11 that calculates weighted average of sixteen pixel values nearest to $\left(\frac{r\prime}{L}, \frac{c\prime}{L}\right)$ coordinates resulting in the interpolated pixel value. Note that the Eq. 11 is formed from Eq. 7, Eq. 8, Eq. 9, and Eq. 10.

$$P(r', c-1) = -a \times (1-a)^2 \times S[r-1, c-1] + (1 - 2 \times a + a^3) \times S[r, c-1] + a \times (1 + a - a^2) \times S[r+1, c-1] - a^2 \times (1-a) \times S[r+2, c-1] \quad (7)$$

$$P(r', c) = -a \times (1-a)^2 \times S[r-1, c] + (1 - 2 \times a + a^3) \times S[r, c] + a \times (1 + a - a^2) \times S[r+1, c] - a^2 \times (1-a) \times S[r+2, c] \quad (8)$$

$$P(r', c+1) = -a \times (1-a)^2 \times S[r-1, c+1] + (1 - 2 \times a + a^3) \times S[r, c+1] + a \times (1 + a - a^2) \times S[r+1, c+1] - a^2 \times (1-a) \times S[r+2, c+1] \quad (9)$$

$$P(r', c+2) = -a \times (1-a)^2 \times S[r-1, c+2] + (1 - 2 \times a + a^3) \times S[r, c+2] + a \times (1 + a - a^2) \times S[r+1, c+2] - a^2 \times (1-a) \times S[r+2, c+2] \quad (10)$$

$$D(r', c') = -b \times (1-b)^2 \times P[r', c-1] + (1 - 2 \times b + b^3) \times P[r', c] + b \times (1 + b - b^2) \times P[r', c+1] - b^2 \times (1-b) \times P[r', c+2] \quad (11)$$

Like NN and BI cases, the Eq. 11 was adapted to B-mode frame from the ultrasound scanner following examples in [19], [20].

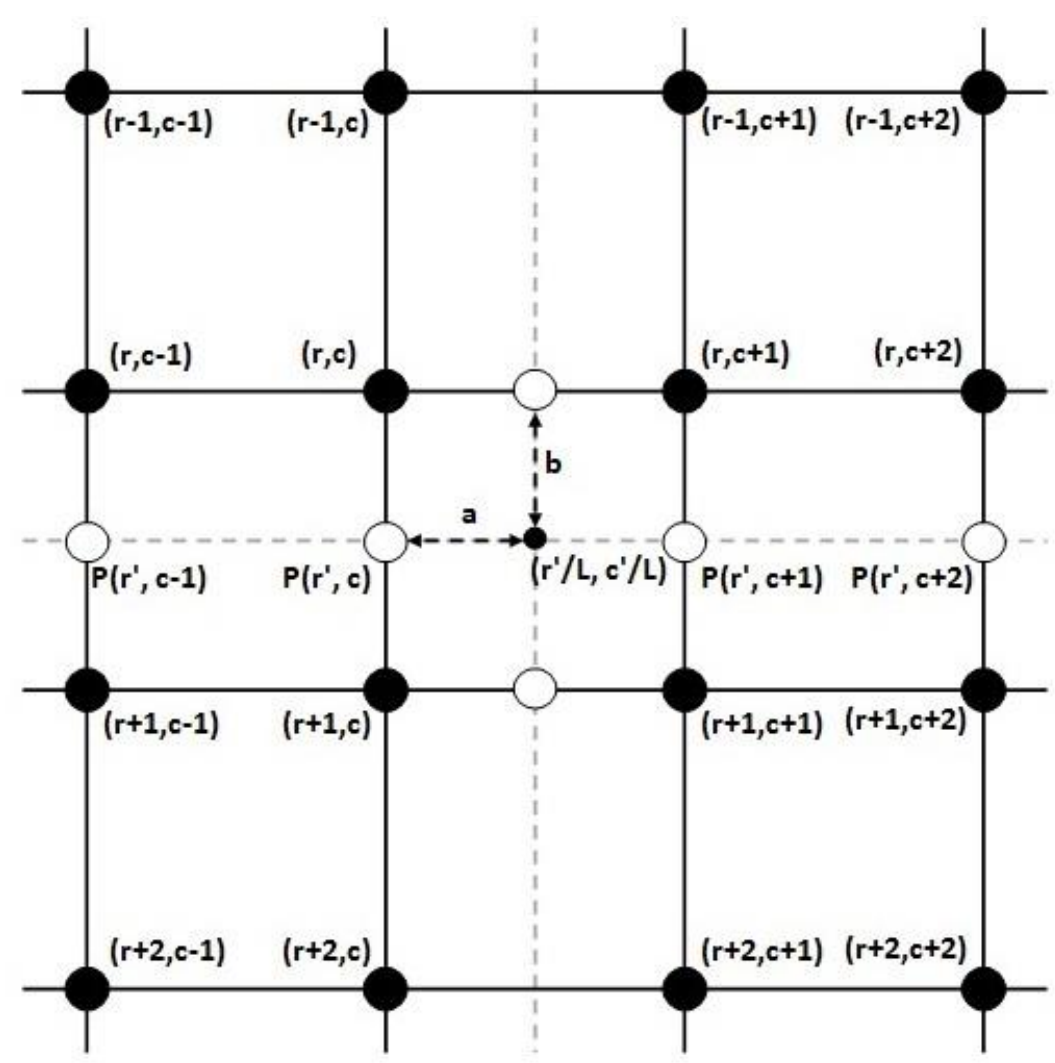

**Figure 5**: Illustration of the pixel grid for BIC image interpolation

### *2.3. Datasets and Metrics*

#### 2.3.1. T5D dataset

T5 Duke University ultrasound scanner or T5D scanner is an ultrasound scanner used to generated data files in which our datasets was created [27]. Our dataset consisted of full sectored images or single video frames obtained by scan-converting only one frame from T5D files, with settings such as Gamma = 0.6, Brightness = 4.2, and Contrast = 120.4. Note that T5D files are not publicly available files. However, upon request full sectored images can be shared with reviewers only for peer-review or clinical relevance evaluation purposes. Examples of scan-converted videos frames are available on this YouTube link, here: https://www.youtube.com/@rukundoo (see: Nearest Neighbor, Bilinear and Bicubic Interpolated Scan Conversion of B-Mode Cardiac Ultrasound Image Sequences – 60).

#### 2.3.2. USC-SIPI dataset

This dataset consisted of ten test images are downloadable from the USC-SIPI Image Database [22]. Downloaded images are: 1) Aerial | 512x512 pixels | 8 bits/pixel | database ref. 5.2.09. 2) Stream and bridge | 512x512 pixels | 8 bits/pixel | database ref. 5.2.10. 3) Fishing Boat | 512x512 pixels | 8 bits/pixel | database ref. boat.512. 4) Male | 1024x1024 pixels | 8 bits/pixel | database ref. 5.3.01. 5) Couple (NTSC test image) | 512x512 pixels | 8 bits/pixel | database ref. 5.2.08. 6) house | 512x512 pixels | 24 bits/pixel | database ref. house. 7) Peppers | 512x512 pixels | 24 bits/pixel | database ref. 4.2.07. 8) Sailboat on lake | 512x512 pixels | 24 bits/pixel | database ref. 4.2.06. 9) Airplane (F-16) | 512x512 pixels | 24 bits/pixel | database ref. 4.2.05. 10) Mandrill (a.k.a. Baboon) | 512x512 pixels | 24 bits/pixel | database ref. 4.2.03. Note that all these images were pre-processed to match basic properties before use for performance evaluation purposes.

#### 2.3.3. Full-reference and non-reference Metrics

Full-reference (FR) and non-reference (NR) metrics are two categories of objective image quality evaluation metrics, widely used in digital image processing [6], [7], [23], [24], [25] [26], to automatically assess the quality of digital images. Those selected, in this work, are: 1) *FR*: Mean-Squared Error (MSE), Structural Similarity Index (SSIM), and Peak Signal to Noise Ratio (PSNR). 2) *NR*: Blind/Referenceless Image Spatial Quality Evaluator (BRISQUE), Perception-based Image Quality Evaluator (PIQE), and Naturalness Image Quality Evaluator (NIQE).

### *2.4. Scheme for estimating the image quality-guided coefficient*

2.4.1. Hypothesis

Referring to Figure 4 and Figure 5, the variables $a$ and $b$ represent the pixel coordinates in x- and y-directions in the destination image. These are mathematically expressed as shown in Eq. 12, where $x = r'/L$, $y = c'/L$ and $\lfloor ... \rfloor$ is a floor rounding function useful when x and y values are non-integers.

$$a = x - \lfloor x \rfloor$$
$$b = y - \lfloor y \rfloor \quad (12)$$

Assume that $k$ is a scalar value carrying general information obtained from evaluating the quality of different interpolated images. Here, we agree that there exist possibly many ways to develop and apply such a coefficient $k$ in the efforts to optimize image interpolation outcomes. Now, given the need for not sacrificing the efficiency of an interpolation algorithm (which is another important criterion while selecting an image interpolation algorithm), we simply multiply the scalar coefficient $k$ to Eq. 12 , as shown in Eq. 13, in the effort to update the BIC's sixteen coefficients (mathematically determined without attempting to adapt them to the reality or complexity of digital image quality).

$$a = k \times (x - \lfloor x \rfloor)$$
$$b = k \times (y - \lfloor y \rfloor) \quad (13)$$

2.4.2. Estimating the coefficient $k$

To estimate the coefficient $k$, two possibilities were considered, namely: (a) the use of feature selection through line chart data visualization on the real number line and (b) the use the objective image quality assessment (IQA) metrics. In the real number set, $\mathcal{R}$, the real number line is a line that extends infinitely in both directions with zero at mid-point. In this context, it is not practical to test all real values on Eq. 13. Therefore, in the search for the value or coefficient $k$, that would lead to the best IQA-based scores; ab initio we used (a) to select the most relevant and smallest point-interval of useful real values. Ab secondo, we found that line to be between $-3$ and $+3$ as shown in Figure 7. Here, it is important to note that [17] proved that the visualization technique was at least as powerful as the best automatic feature selection algorithms. Figure 6 illustrates key steps needed for estimation of the coefficient-$k$.

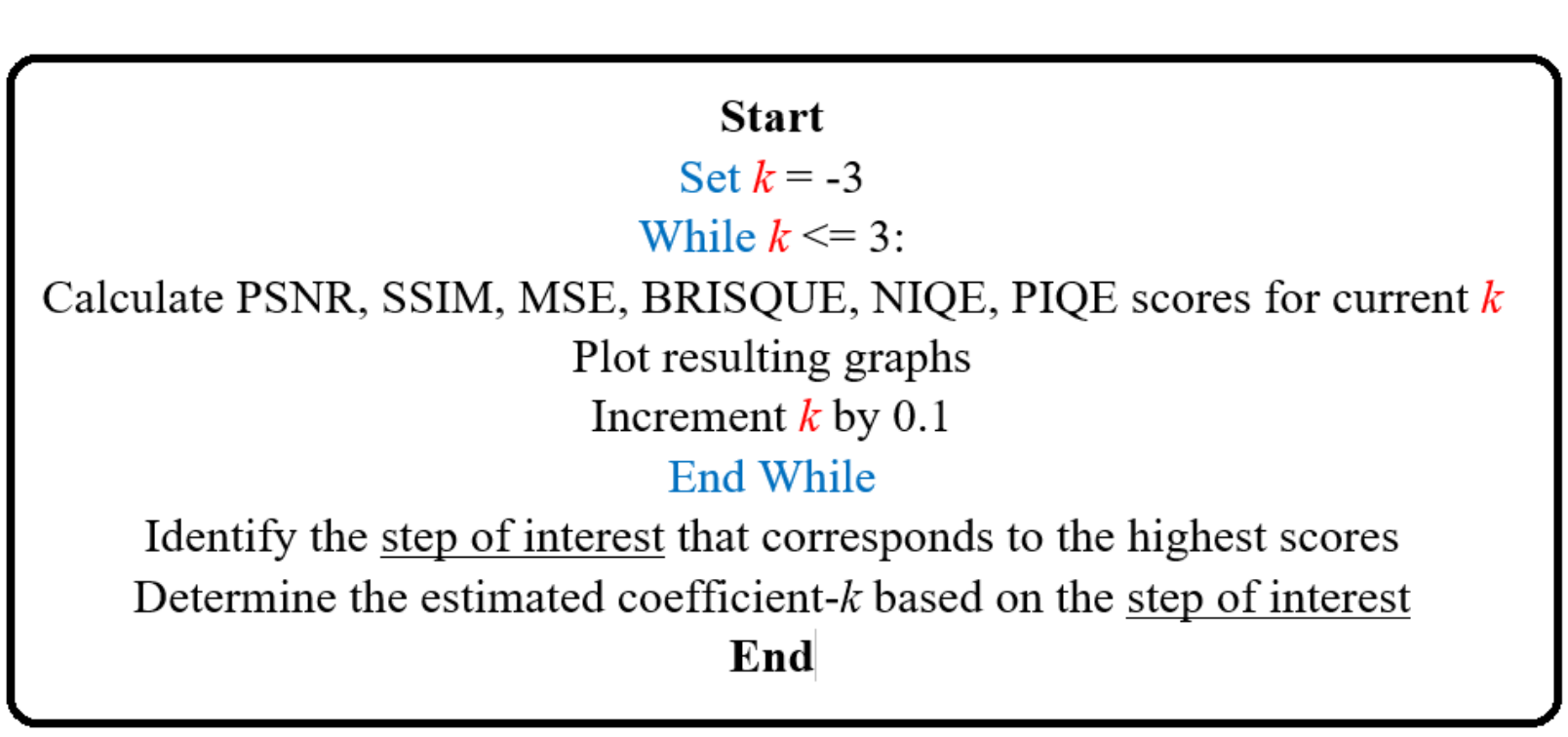

**Figure 6**: Key steps for estimating the coefficient-$k$

To estimate the best coefficient $k$ using Eq. 13, $k$ is exposed to all real values between $-3$ and $+3$ with a 0.1 step, as shown in Figure 6. At each step, we assess the image quality using metrics mentioned in Figure 6 and resulting graphs are shown in Figure 7. After completing all steps and evaluating all metrics scores, we identify the step matching with or corresponding to (the majority of) the best IQA scores to determine the estimated coefficient-$k$ of interest. Figure 8 shows the average value of the coefficient k (1.1833) estimated from ten images that were upscaled at different scaling factors, namely 2X, 4X, and 8X.

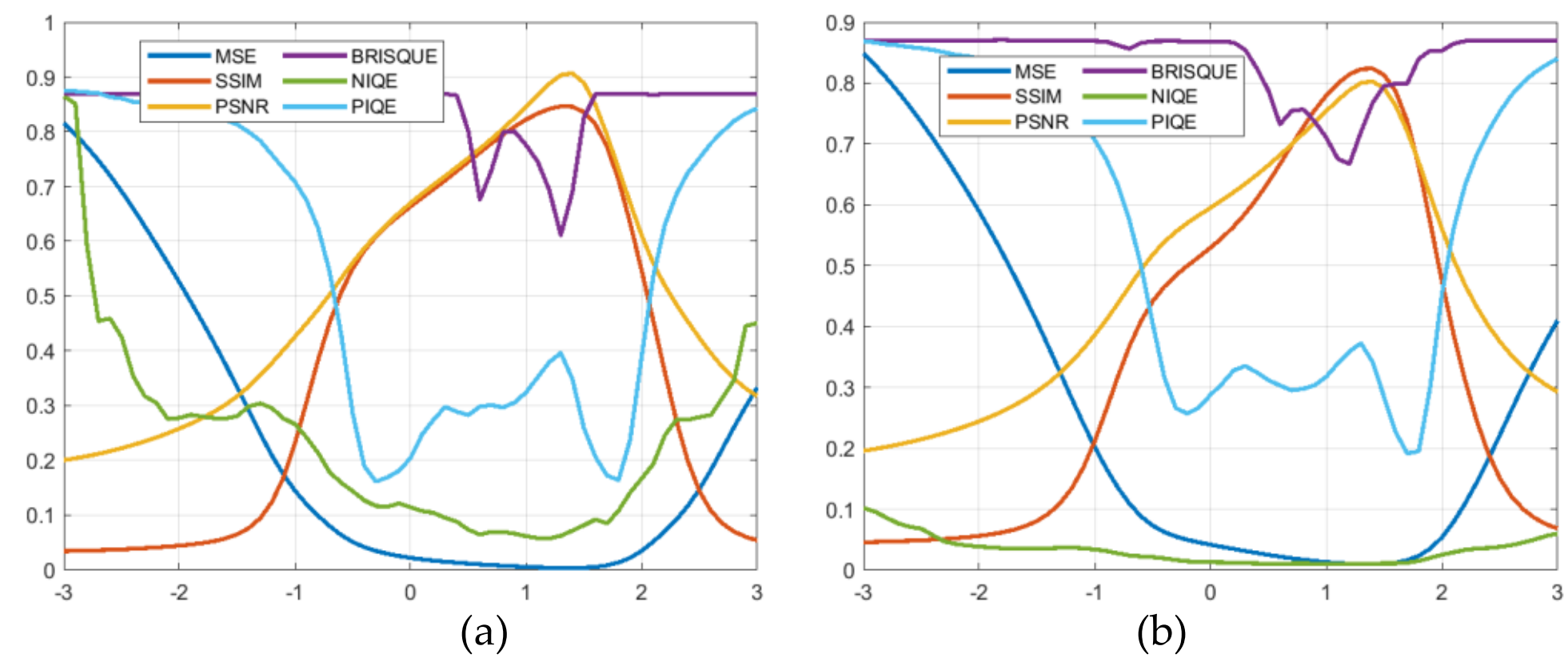

**Figure 7**: (a) Graph for 1 image. (b) Graph for averaged 10 images (see: Fig.8).

Note that Figure 7-(a) and Figure 7-(b) prove that each image can have a unique coefficient *k*, as averaged ten images' graph (a) does not look exactly like a single image graph (b). Also, note that according to our hypothesis, we expected the effects of the coefficient *k* - determined from ground truth natural images (since there are no ground truth ultrasound images) - on natural images would be almost the same as on ultrasound sectored images.

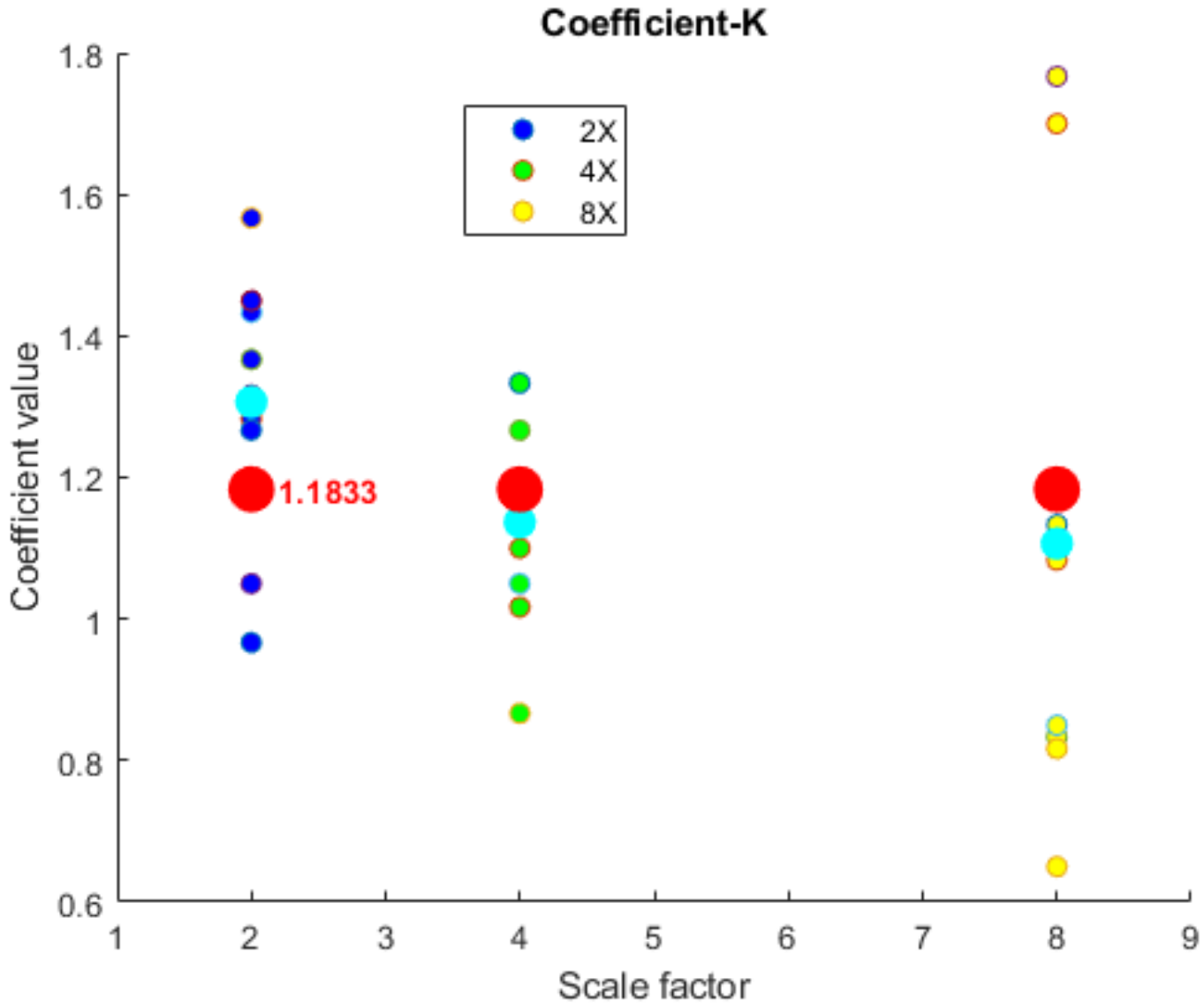

**Figure 8**: Estimated coefficient *k*= *1.1833*

Note that the multiplication of a scalar coefficient *k* by the BIC coefficients cannot affect significantly the efficiency of the BIC algorithm. Therefore, in this paper, it would not be meaningful to evaluate the OBIC against the BIC algorithms in terms of speed of execution.

## 3. Results and Discussions

### *3.1. Performance evaluation on natural images*

Objective evaluation of the performance of NN, BI, BIC, Lanczos-2 kernel image interpolation (LA2), Lanczos-3 kernel image interpolation (LA3), and OBIC algorithms is presented in Figures 9, Figure 10, and Figure 11. Here, the focus was on objective evaluation, not only because we wanted to provide a more rigorous and quantitative analysis of the algorithms' performance, which could be replicated by other researchers, but also due to availability of reliable ground truth images.

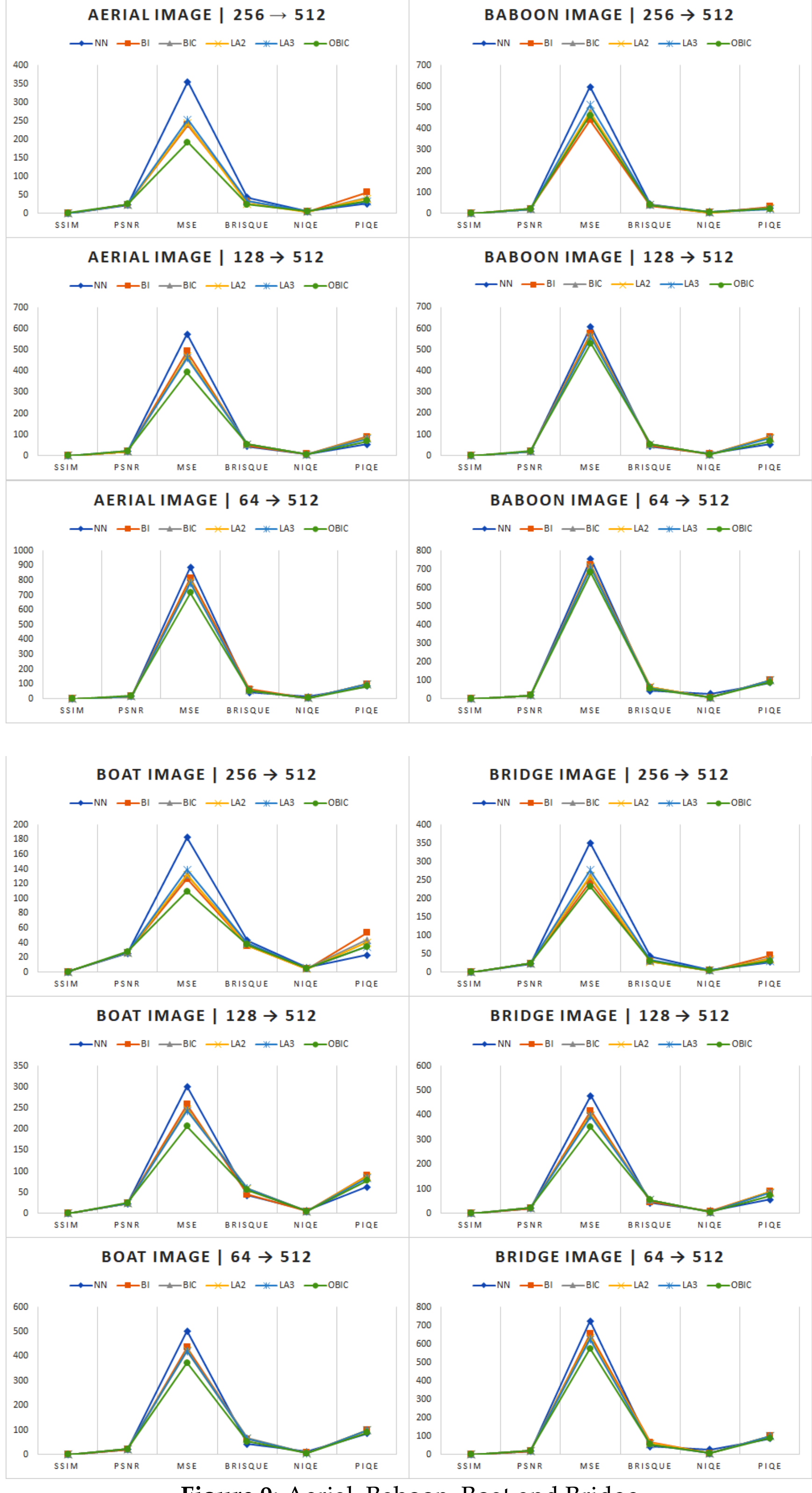

**Figure 9:** Aerial, Baboon, Boat and Bridge

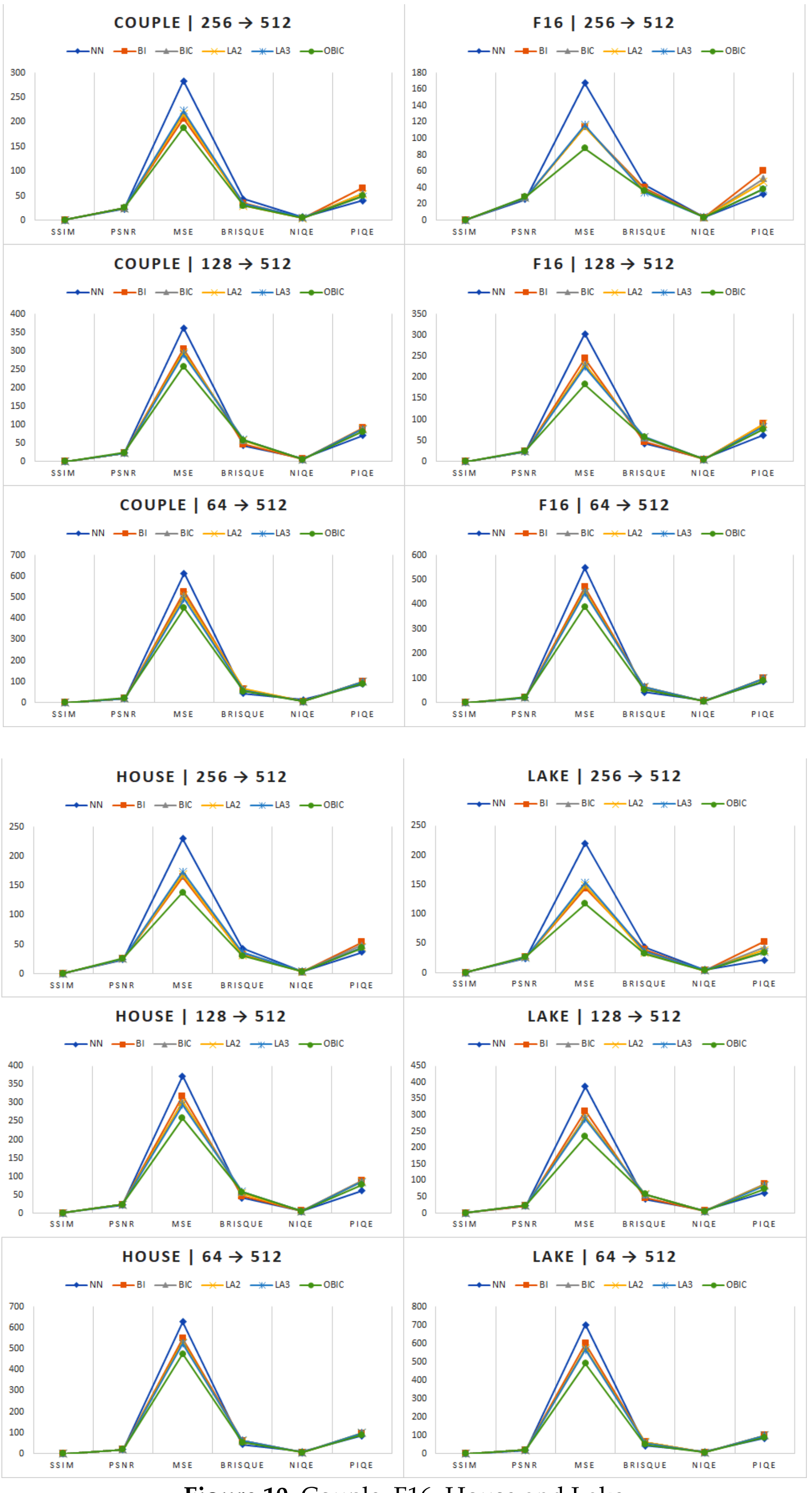

**Figure 10**: Couple, F16, House and Lake

Now, referring to the algorithm's performance evaluation scheme developed in [16], in Figures 9, Figure 10, and Figure 11 we have six interpolation algorithms evaluated on ten images using six IQA metrics at three different scaling ratios or factors. Understandably this means that each one of six interpolation algorithms has chance for a maximum occurrence frequency equals to 180 occurrences (i.e. 6 metrics × 3 ratios × 10 images). In other words, each algorithm can outperform others or achieve the best IQA metric-based scores 180 times. In this context, we considered two cases, namely: a) General case and b) Specific case.

(a) - Generally, evaluating OBIC algorithm against other algorithms mentioned, the OBIC achieved 97.78 % of 90 occurrences, only using FR IQA metrics. In other words, the OBIC algorithm performed better than all methods mentioned 88 times over the expected 90 times (i.e., excluding NR metrics). In the group of NR IQA metrics alone, the OBIC algorithm achieved 16.67 % of 90 occurrences. This means that the OBIC algorithm performed better than all other methods 15 times over the expected 90 times (which also means that it was not the most highly ranked by NR IQA metrics).

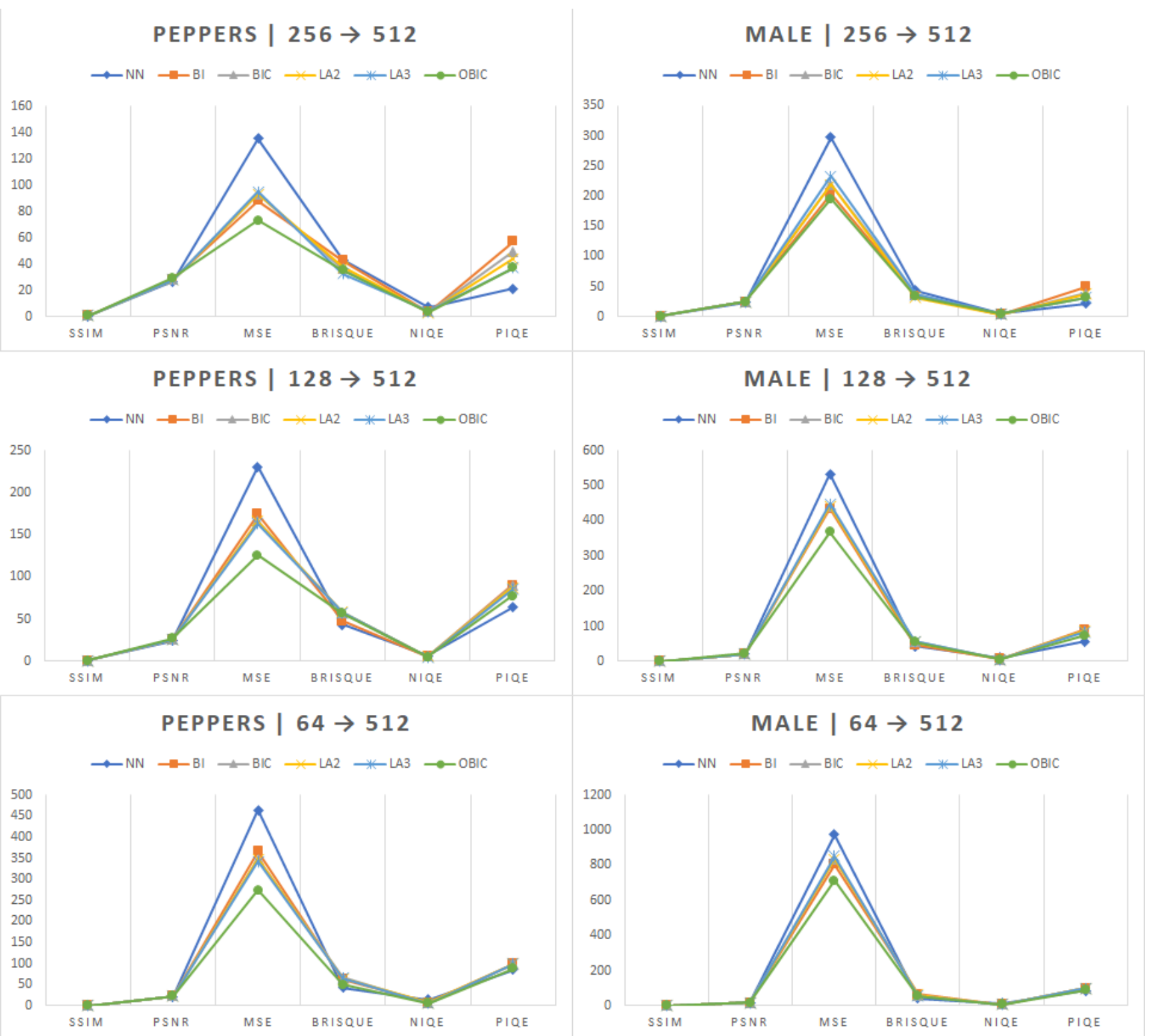

**Figure 11**: Peppers and Male

Here, the overall OBIC performance was equivalent to 57.22% of 180 occurrences, which means that it outperformed other methods 103 times over the expected 180 times.

(b)-Specifically, evaluating the OBIC against the BIC algorithm: In the group of only FR IQA metrics, the OBIC achieved 100% of 90 occurrences. In other words, the OBIC algorithm performed better than the BIC method 90 times over the expected 90 times. In the group of NR IQA metrics, the OBIC algorithm achieved 84.44% of 90 occurrences. This means that the OBIC algorithm performed better than the BIC method 76 times over the expected 90 times. Here, the overall performance of the OBIC method was equivalent to 92.22% of 180 occurrences, which means that it outperformed the BIC method 166 times over the expected 180 times.

### *3.3. Performance evaluation on ultrasound images*

In case of ultrasound images, where ground truth images did not exist, we only subjectively evaluated the performance of the resulting interpolated scan conversion algorithms. In this context, we used cropped whole sectored images to assess the quality of cropped images or frames. Here, criteria to assess the perceptual image quality was centered on crispness and/or contrast of edge regions because of visual artefacts occurring frequently in such regions. In this context, image crispness can be understood the sharpness or clarity of details in an image while image contract can be understood as the difference in image details that make an object or objects distinguishable.

Ab initio, we generally evaluated the perceptual quality of top two and one bottom left cropped sectored images generated by traditional NN, BI and BIC interpolated scan conversion algorithms.

At the top left, the NN interpolated scan conversion algorithm produced the only images with the crispest edges (see Figure 12, Figure 13 and Figure 14). Despite that, its overall quality remained poor due to the presence of jaggies that do not allow more details to be seen.

At the top right, the BI interpolated scan conversion algorithm produced images that generally looked better than in the previous NN cases, especially, around the oblique edge (see Figure 12, Figure 13 and Figure 14). However, it could be seen that the perceived blurriness artefacts reduced the overall quality in terms of contrast and crispness.

At the bottom left, the BIC interpolated scan conversion algorithm produced images looking better than in the previous two cases of NN and BI, only in terms of contrast (see Figure 12, Figure 13 and Figure 14). Despite improvements in terms of visual or perceptual quality, BIC algorithm slightly produced blurring artifacts as well as rhombic spots (to some extent) on the oblique edges.

Ab secondo, the quality of the bottom two cropped sectored images from scan conversion algorithms generated by BIC and OBIC interpolated scan conversion algorithms were particularly evaluated for each Figure's case.

In Figure 12, the OBIC interpolated scan conversion algorithm produced an image with an oblique edge that looked crisper than that of the BIC interpolated scan conversion algorithm. With the BIC interpolated scan conversion algorithm, there were visible blurriness and ringing artefacts in the rhombic highlight spot areas. But, in the OBIC image case, there is no rhombus like areas due to producing clearly defined details that lead to better visual contrast than in the BIC case.

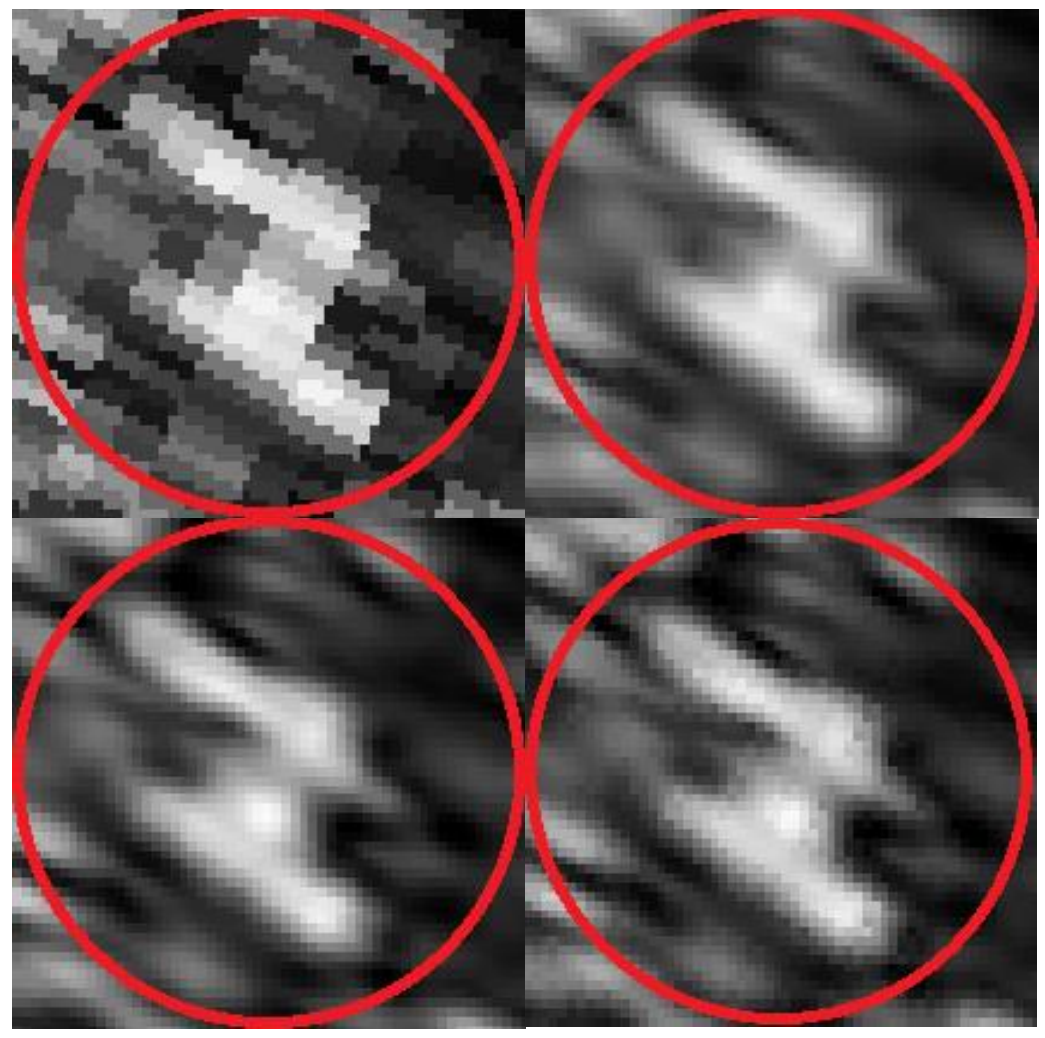

**Figure 12**: *Cropped image 1* - evaluation of oblique edges. From top-left to top-right, NN-, BI- and bottom-left to bottom-right, BIC- and OBIC interpolated scan conversion algorithms output cropped images.

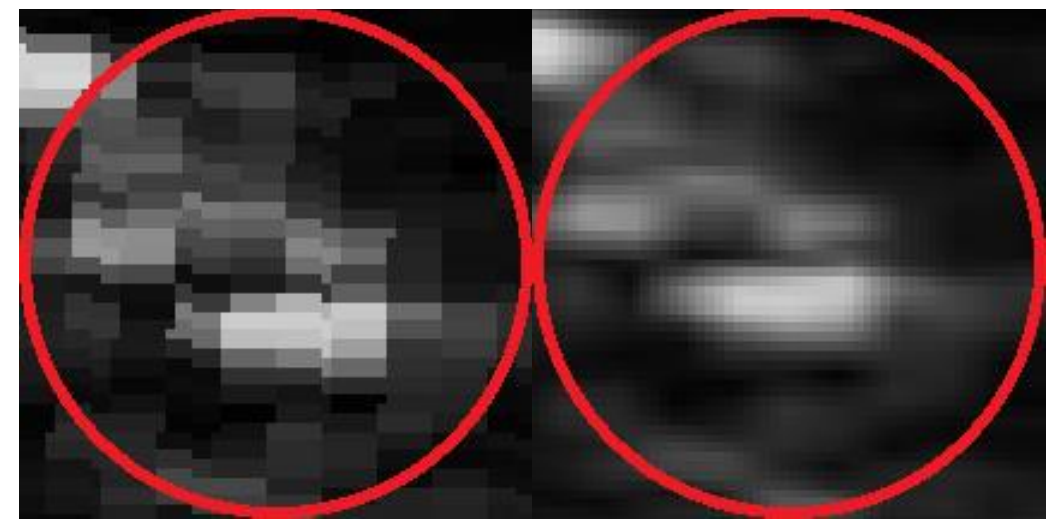

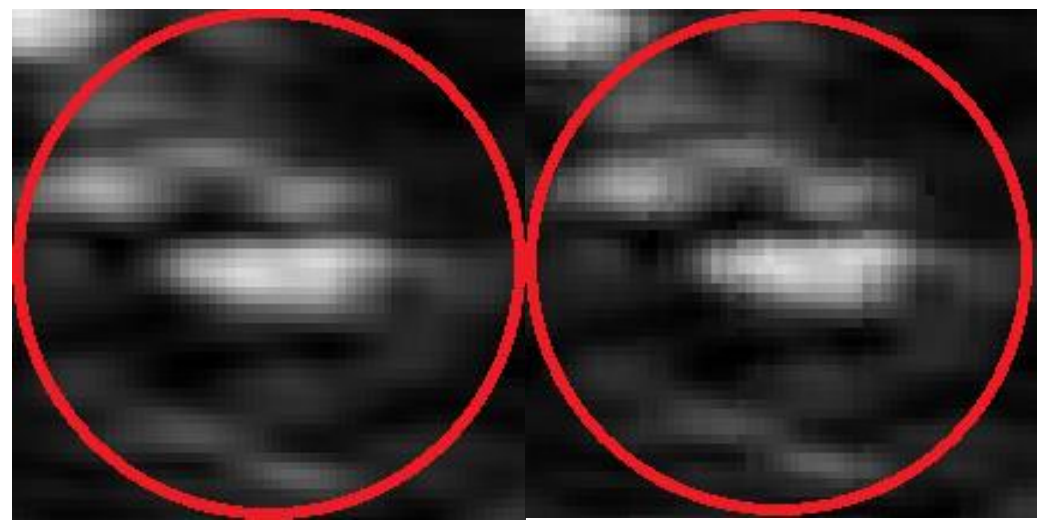

**Figure 13**: *Cropped image 2* - evaluation of horizontal edges. From top-left to top-right, NN-, BI- and bottom-left to bottom-right, BIC- and OBIC interpolated scan conversion algorithms output cropped images.

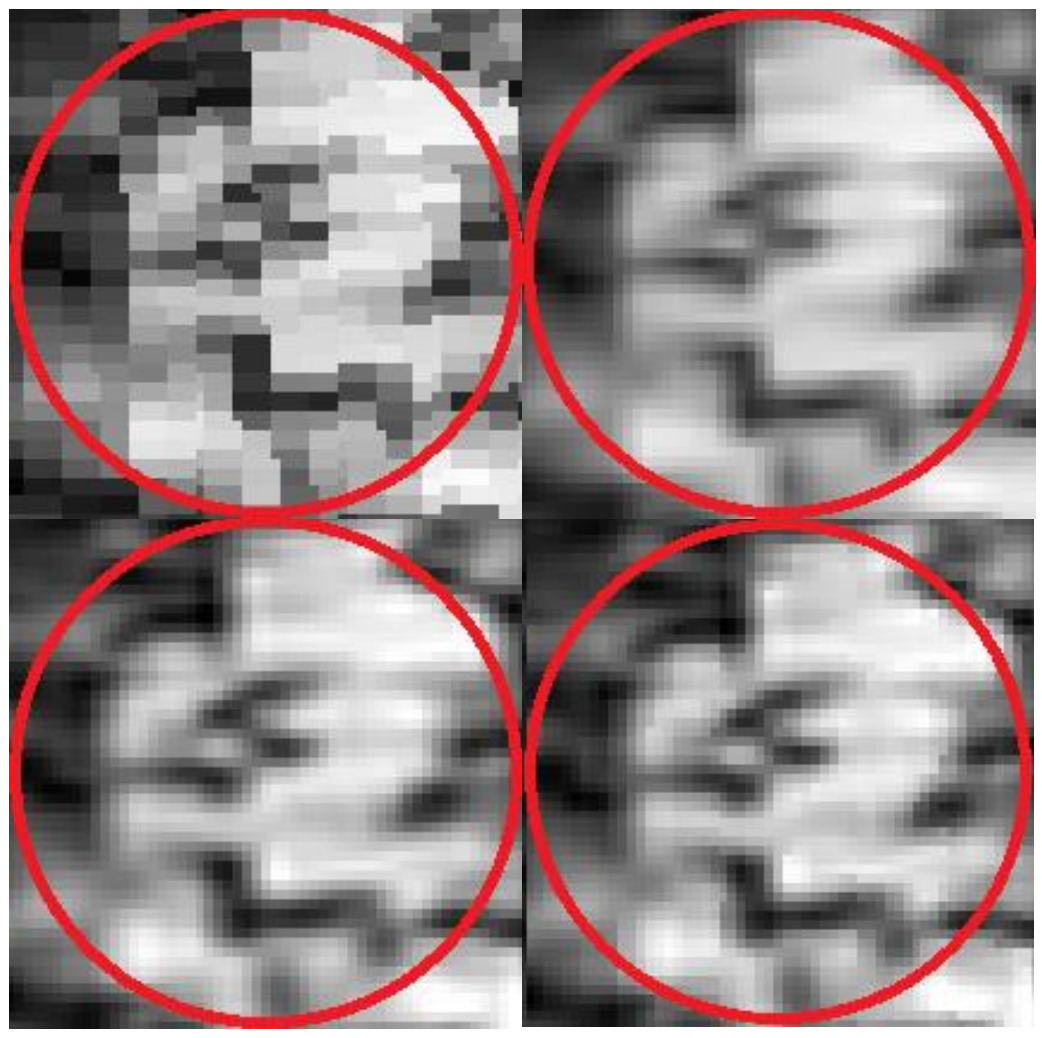

**Figure 14**: *Cropped image 3* - evaluation of vertical edges. From top-left to top-right, NN-, BI- and bottom-left to bottom-right, BIC- and OBIC interpolated scan conversion algorithms output cropped images.

In Figure 13, on the other hand, the horizontal edge in the OBIC image became much crisper than that of the BIC image. As can be seen, the edge in the BIC image is surrounded by square blurriness like details thus making it difficult to see the boundary of the edge, which is not the case with the edge in the OBIC image. Also, the OBIC image has better overall contrast and visibility than in the BIC image. Some of the details in the BIC faded while remained visible in the OBIC image.

In Figure 14, there are complex edge structures. With the BIC image, the vertical edge lines look almost the same way as in the OBIC image case. The slight difference is the level of contrast since, in the OBIC image case, the black and white regions are clearer than in the BIC image case. In general, the OBIC interpolated scan conversion algorithm produced crisper and better contrast cropped sectored images than the BIC algorithm, as well as other interpolated scan conversion algorithms, mentioned.

It is important to note that perceptual differences between these images are more visible on full screens than in print or this paper. It can be understood that crispness and contrast are preferred in this context, as the opposite would generally result in blurry artifacts, reducing the edge and texture details of sectored images, which would have a direct consequence on the variability of measurements in medical diagnosis.

## 4. Conclusions

The optimized bicubic (OBIC) algorithm developed using an image quality assessment (IQA) metrics-based strategy has demonstrated superior performance over the traditional bicubic (BIC) algorithm. Experimental simulations have shown that the IQA metrics-based strategy significantly improves the performance of the BIC algorithm, resulting in an overall occurrence-based improvement of 92.22%. Also, experiments revealed the need for improving the quality of edge and texture details in ultrasound sectored images taking into account that interpolated scan conversion image quality cannot be adjusted during the patient's examination. Further research may focus on making image interpolation algorithms capable of producing realistic estimates as OBIC algorithm only performed limitedly when assessed using non-reference image quality assessment (NR-IQA) metrics.

**Author Contributions:** Conceptualization, methodology, software or algorithm development, validation, formal analysis, investigation, original draft preparation, review and editing of the manuscript, OR.; resources, data curation and supervision, SES.; project management and funding acquisition, OTVR.

**Funding:** The research project was funded by Novo Nordisk Foundation/Novo Seed. The research work was supported by Aalborg and Duke Universities between March and December 2018.

**Institutional Review Board Statement:** Not applicable.

**Informed Consent Statement:** Not applicable.
**Data Availability Statement:** Data are available on our servers and can be shared upon request.

**Acknowledgments:** Authors would like to thank researchers from the Cardio Tech Research Group and Von Ramm Ultrasound Laboratory at Aalborg and Duke Universities, respectively for their support.

**Conflicts of Interest:** The authors declare no conflict of interest.